\documentclass[aps,prb,twocolumn,groupedaddress,showpacs]{revtex4}
\usepackage{color}
\usepackage{booktabs}
\usepackage{graphicx}
\usepackage{epsfig}
\usepackage{amssymb}
\usepackage{bm}
\usepackage{graphpap}
\usepackage[normalem]{ulem}
\usepackage[dvipdfmx]{hyperref}
\usepackage[all]{hypcap}
\usepackage{url}
\hypersetup{colorlinks=true,citecolor=blue,linkcolor=red}

\begin{document}
\newdimen\heavyrulewidth
\newdimen\lightrulewidth
\newdimen\cmidrulewidth
\newdimen\belowrulesep
\newdimen\belowbottomsep
\newdimen\aboverulesep
\newdimen\abovetopsep
\newdimen\defaultaddspace
\heavyrulewidth=.08em
\lightrulewidth=.05em
\belowrulesep=.65ex
\belowbottomsep=0pt
\aboverulesep=.4ex
\abovetopsep=0pt
\defaultaddspace=.5em

\title{Size, oxidation, and strain in small Si/SiO$_2$ nanocrystals.}

\author{Roberto \surname{Guerra}}
\affiliation{CNR-INFM-$S{^3}$ and Dipartimento di Fisica -
Universit\`a di Modena e Reggio Emilia - via Campi 213/A I-41100
Modena Italy.}
\author{Elena \surname{Degoli} and Stefano \surname{Ossicini}}
\affiliation{CNR-INFM-$S{^3}$ and Dipartimento di Scienze e Metodi
dell'Ingegneria - Universit\`a di Modena e Reggio Emilia - via
Amendola 2 Pad. Morselli, I-42100 Reggio Emilia Italy.}
\date{\today}

\begin{abstract}
The structural, electronic and optical properties of Si nanocrystals of different size and shape, passivated with hydrogens, OH groups, or embedded in a SiO$_2$ matrix are studied. The comparison between the embedded and free, suspended nanocrystals shows that the silica matrix produces a strain on the embedded NCs, that contributes to determine the band gap value. By including the strain on the hydroxided nanocrystals we are able to reproduce the electronic and optical properties of the full Si/SiO$_2$ systems. Moreover we found that, while the quantum confinement dominates in the hydrogenated nanocrystals of all sizes, the behaviour of hydroxided and embedded nanocrystals strongly depends on the interface oxidation degree, in particular for diameters below 2 nm. Here, the proportion of NC atoms at the Si/SiO$_2$ interface becomes relevant, producing surface-related states that may affect the quantum confinement appearing as inner band gap states and then drastically changing the optical response of the system.
\end{abstract}

\maketitle

\begin{figure*}[ht]\begin{center}
\centerline{
    \psfig{file=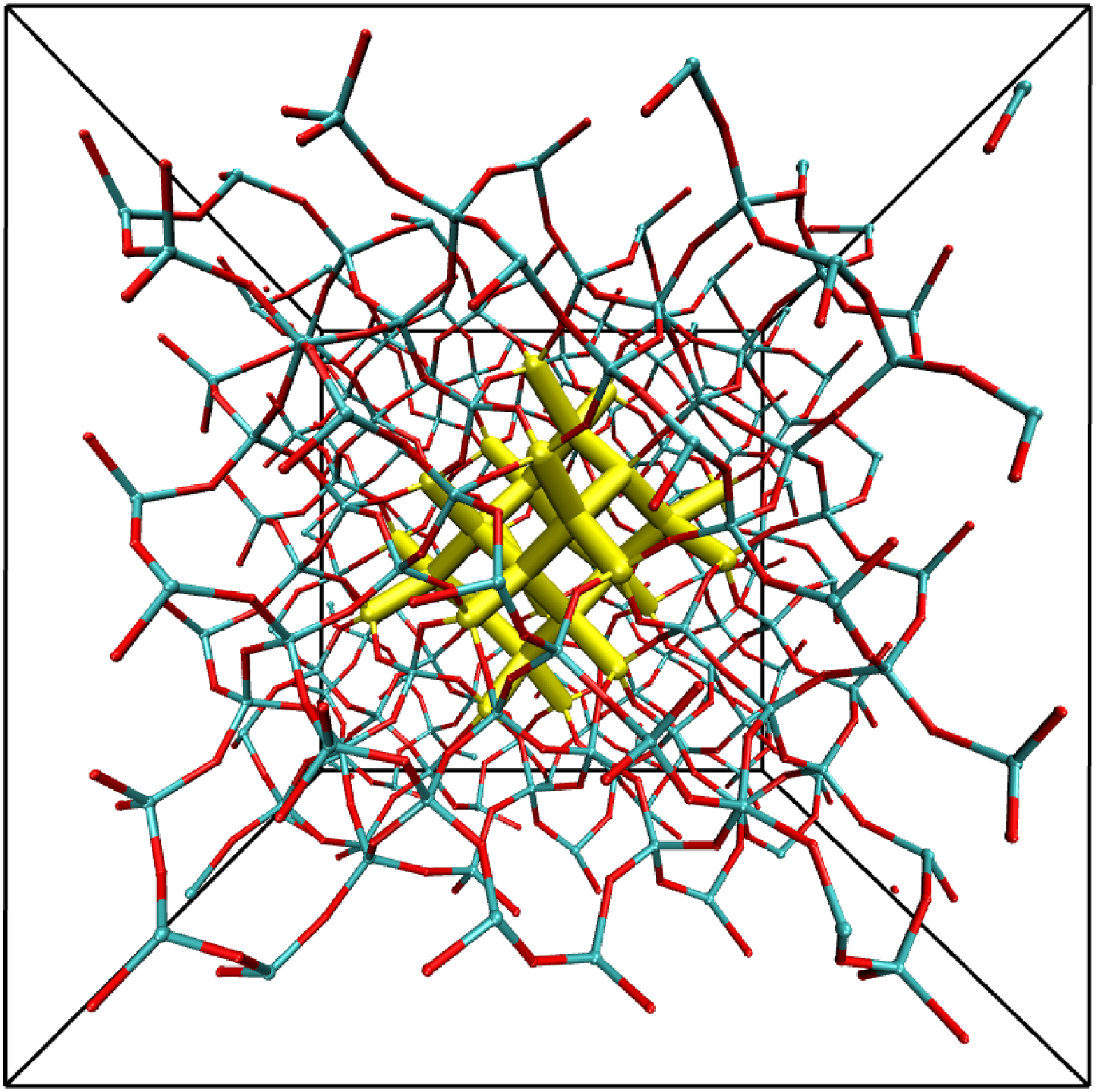,width=5.5cm}~
    \psfig{file=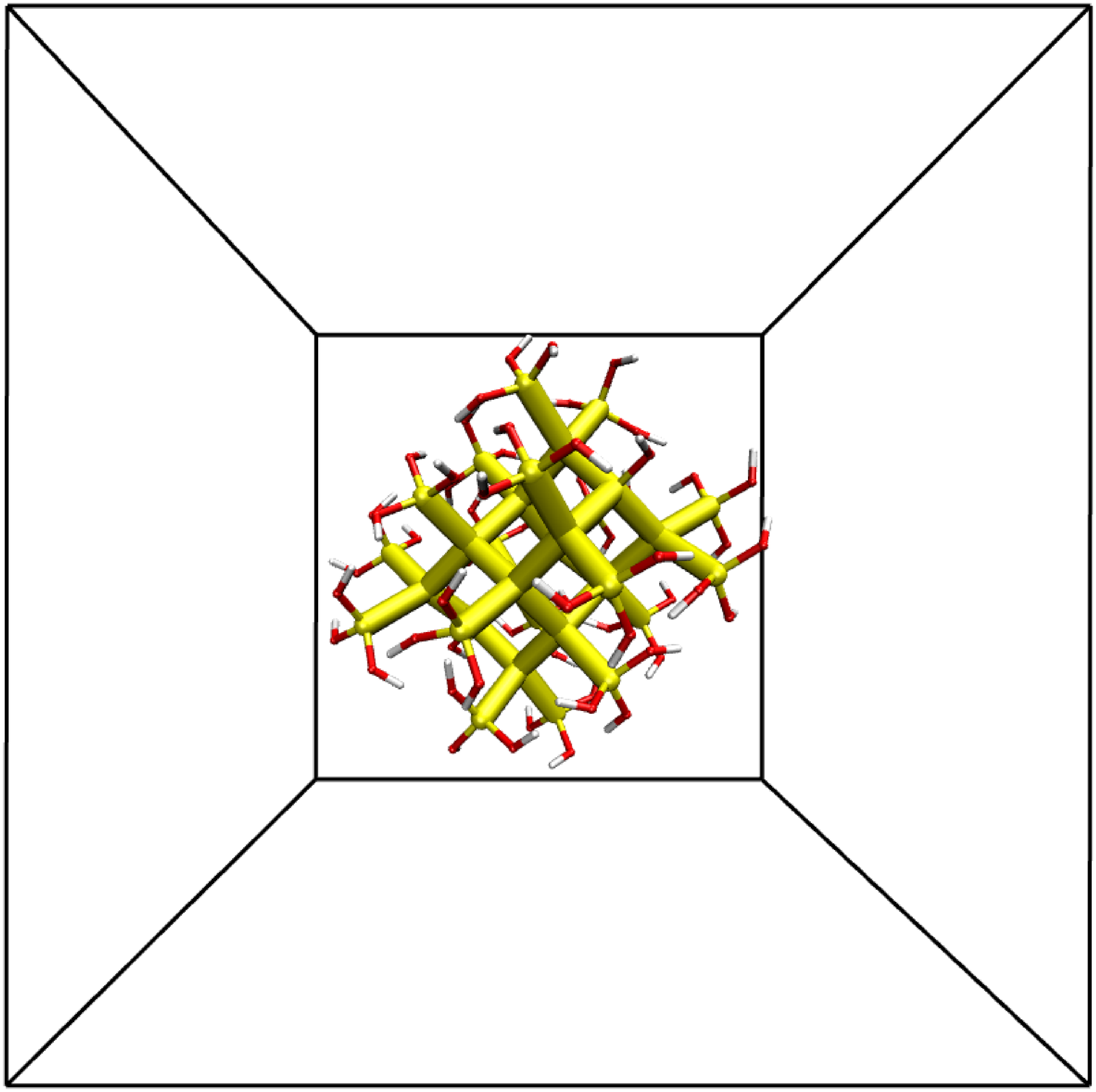,width=5.5cm}~
    \psfig{file=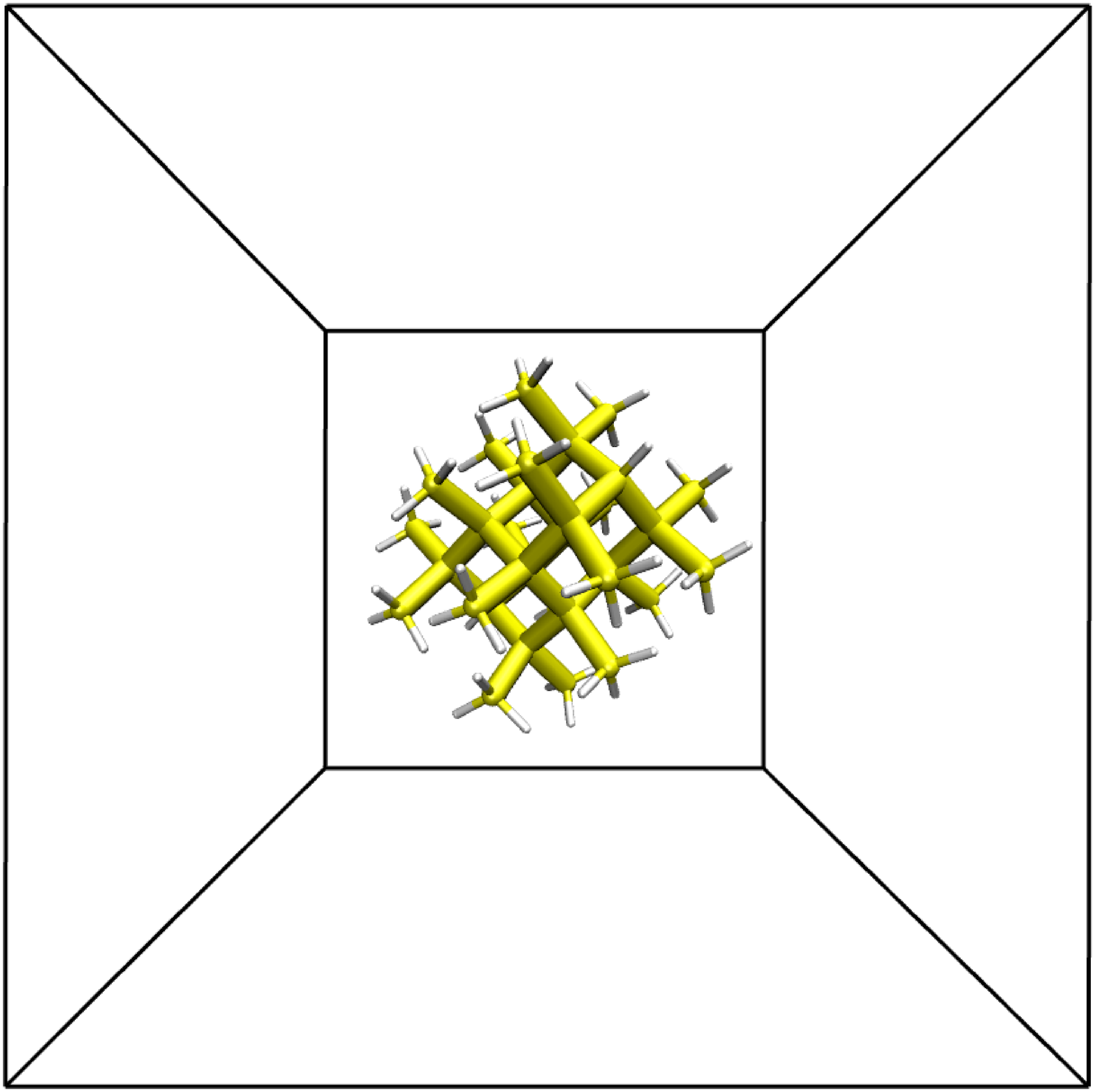,width=5.5cm}}
\caption{(color online) The final optimized structure of the Si$_{32}$ in $\beta$-cristobalite matrix (left panel), Si$_{32}$-OH (center panel), and Si$_{32}$-H (right panel). Red (dark gray) sticks represent the O atoms, cyan (gray) sticks represent the Si of the matrix, and the yellow (gray) thick sticks represent the Si atoms of the NC.}
\label{struttura}\end{center}\end{figure*}

\section{Introduction}
Silicon nanocrystals (Si-NCs) have known very intensive research activity since last decade because of their very promising light emitting properties.\cite{bisi,pavlockwood} The confinement of carriers in silicon structures leads to a strong enhancement of the radiative transition yield and to increase of the emitted photon energy. Embedding Si-NCs in wide band gap insulators is one way to obtain a strong quantum confinement (QC). Si-NCs embedded in a silica matrix have been obtained with several techniques as ion implantation,\cite{brongersma,iwayama} chemical vapour deposition,\cite{hernandez,rolver,godefroo,gardelis} laser pyrolysis,\cite{delerue,kanemitsu} electron beam lithography,\cite{sychugov_linnros} sputtering,\cite{averboukh,antonova} and some others. Alternative preparation methods produces NCs with different surface terminations (D, F, C, and others) that earned some interest in the last years.\cite{salivati,liptak,coxon,konig,lehtonen} Moreover, also the role of nitrogen has been recently explored\cite{dalnegro}.

If the response of single NCs can be described as atomic-like,\cite{sychugov_linnros,empedocles_norris_bawendi,delerue} in experimental samples no two NCs are the same, and collective properties have to be measured. Experimentally, several factors contribute to make the interpretation of measurements on these systems a difficult task. For instance, samples show a strong dispersion in the NC size, that is difficult to be determined. In this case it is possible that the observed quantity does not correspond exactly to the mean size but instead to the most responsive NCs.\cite{credo} Again, NCs synthesized by using different techniques often show different properties in size, shape and in the interface structure.
The characterization of the effects of these quantities (size, interface configuration and passivation types) on the final properties of the system, is of fundamental importance in order to determine the best condition for light emission and optical gain.

Previous works\cite{guerra,vasiliev,allan_delerue_lannoo} already observed that, while the electronic gap presents an oscillating behaviour with the size of the NC, the absorption threshold follows the QC for all the size considered. The effect is that the expected photoluminescence (PL) energy (that is associated to the electronic band gap) and the observed Stokes-shift (that is related to the difference between the absorption and emission energies), cannot be simply described as a function of the bare NC size, but the specific configuration of the interface should be taken into account.\cite{kanemitsu}\\
It has also been demonstrated that, for these systems, while the inclusion of local fields provide an important contribution to the absorption spectra, the excitonic and Coulomb corrections almost exactly cancel out each other.\cite{guerra} This allows a simplification of the algorithms adopted in the calculations, still producing reliable results.\\
In this work we present first-principles calculations of the structural, electronic, and optical properties of Si-NCs of different size and shape, passivated with hydrogens, OH groups, and embedded in a SiO$_2$ matrix. The effects generated by the surrounding matrix on the electronic properties of the NCs (of different size) are analyzed. In particular, we compare a set of suspended (free) and embedded NCs in order to investigate the effects of the strain and the role of oxidation on the final properties of the system.\\
The paper is organized as follows. A description of the theoretical methods and of the systems under investigation is given in section \ref{sec_method}. The structural, electronic and optical properties for the embedded and suspended NCs are analyzed in section \ref{sec_embed-vs-susp}. In section \ref{sec_strainox} the effects of the NCs size and oxidation are discussed.
Conclusions are presented in section \ref{sec_conclusions}.

\section{The Method}\label{sec_method}
The $\beta$-cristobalite (BC) SiO$_2$ is well known to give rise to one of the simplest Si/SiO$_2$ interface because of its diamond-like structure.\cite{BC} The crystalline embedded structures have been obtained from a BC cubic matrix by removing all the oxygens included in a cutoff-sphere, whose radius determines the size of the NC. By centering the cutoff-sphere on one silicon or in an interstitial position it is possible to obtain structures with different symmetries. To guarantee a proper shielding of the introduced strain by the surrounding silica we preserved a separation of about 1 nm between the NCs replica.\cite{mluppi1,mluppi2,flyura} The result is a structure with the silicon atoms bonded together to form a crystalline skeleton with T$_d$ local symmetry before relaxation. In such core, Si atoms show a larger bond length (3.1 \AA) with respect to that of the Si bulk structure (2.35 \AA). No defects (dangling bonds) are present, and all the O atoms at the Si/SiO$_2$ interface are single bonded with the Si atoms of the NC.\\
The optimized structure has been achieved by relaxing the total volume of the cell (see example in Fig. \ref{struttura}, left panel). The relaxation of all the structures have been performed using the SIESTA code\cite{siesta1,siesta2} and Troullier-Martins pseudopotentials with non-linear core corrections. A cutoff of $150$ Ry on the density and no additional external pressure or stress were applied. Atomic positions and cell parameters have been left totally free to move.\\
The hydroxided-strained NCs (s-Si$_{xx}$-OH) are obtained by extracting the NCs together with the first interface oxygens from the relaxed NC-silica complexes, and then passivating the surface with hydrogen atoms (Fig. \ref{struttura}, center panel). Also the case of hydrogenated-strained NCs is considered (s-Si$_{xx}$-H), by replacing the OH groups with hydrogens (Fig. \ref{struttura}, right panel). In the last two cases, in order to preserve the strain, only the hydrogen atoms have been relaxed. The equivalent hydroxided-relaxed (r-Si$_{xx}$-OH) and hydrogenated-relaxed (r-Si$_{xx}$-H) NCs are obtained by a further optimization and relaxation of the whole systems.\\
The goal is to distinguish between the properties that depend only on the NC from those that are instead influenced by the presence of the matrix. The comparison of the results relative to different passivation regimes (H or OH groups) could give some insight on the role played by the interface region. Electronic and optical properties of the relaxed structures have been obtained in the framework of the density functional theory (DFT), using the ESPRESSO package.\cite{espresso} Calculations have been performed using norm-conserving pseudopotentials within the LDA approximation with a Ceperley-Alder exchange-correlation potential, as parametrized by Perdew-Zunger. An energy cutoff of 60 Ry on the plane wave basis have been considered.

\section{Embedded VS Suspended NCs}\label{sec_embed-vs-susp}

\begin{figure}[ht]\begin{center}
\centerline{\psfig{file=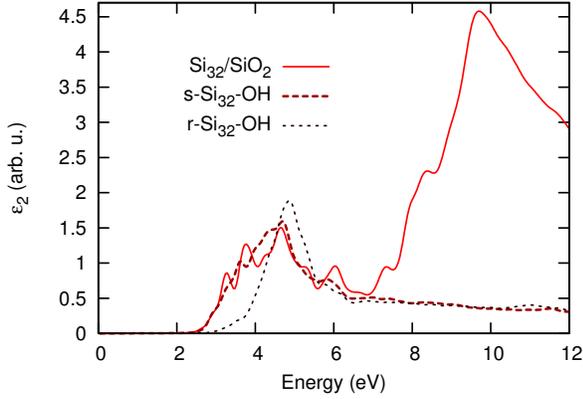,width=5.5cm,angle=270}}
\caption{(color online) DFT-RPA imaginary part of the dielectric function for the Si$_{32}$/SiO$_2$ (solid), s-Si$_{32}$-OH (dashed), r-Si$_{32}$-OH (dotted).}
\label{eps2_strain}\end{center}\end{figure}

In this section we analyze the role of the surrounding silica matrix on the electronic and optical properties of the Si/SiO$_2$ systems. We compared the results of the composite Si/SiO$_2$ systems and the suspended ones, including the effects of the strain induced by the silica matrix.

\begin{table}[b]\begin{center}\begin{tabular}[t]{c @{\hspace{0.6cm}} c
@{\hspace{0.3cm}} c @{\hspace{0.3cm}} c @{\hspace{0.3cm}} c}
\toprule
                      & Si$_{10}$   & Si$_{17}$ & Si$_{32}$ & Si$_{35}$ \\
\midrule
    embedded              & 1.77    & 2.36      & 2.62      & 1.62 \\
    hydroxided-strained   & 1.60    & 2.21      & 2.45      & 1.43 \\
    hydroxided-relaxed    & 1.97    & 3.50      & 2.93      & 1.50 \\
    hydrogenated-strained & 4.66    & 2.51      & 2.90      & 3.14 \\
    hydrogenated-relaxed  & 4.61    & 4.06      & 3.54      & 3.43 \\
\bottomrule
\end{tabular}
\caption{HOMO-LUMO gaps for the embedded and suspended set of NCs. Units are in eV.}
\label{tabella1}\end{center}\end{table}

Four embedded system have been produced: the Si$_{10}$ and Si$_{17}$ structures have been obtained from a BC-2x2x2 supercell (192 atoms), while for the Si$_{32}$ and Si$_{35}$ NCs the larger BC-3x3x3 supercell (648 atoms) has been used.\\
Being the bonds near the interface much more strained with respect to those in the NC-core,\cite{bulutay,flyura,kroll} the average strain tends to zero for large clusters. Therefore we adopted the most strained bond value (MSB) (calculated from the difference of the largest Si-Si distance of the NC, with respect to the bulk value), as the parameter representing the NC strain. For the four structures enumerated above we obtained MSB values ranging between 0.1 and 0.5 \AA. The corresponding suspended, hydroxided and hydrogenated, strained and relaxed NCs were obtained following the procedure described in the previous section.\\
The calculated energy differences between the highest-occupied-molecular-orbital (HOMO) and the lowest-unoccupied-molecular-orbital (LUMO) for all the systems are presented in Table \ref{tabella1}. First of all, we note that the gap of the embedded NCs presents an oscillating behaviour with the NC size (disregarding the QC rule), as already evidenced in other works.\cite{guerra,koponen} For the hydroxided systems we note that, while the removal of the embedding medium produces a constant reduction of the gap of about 0.15-0.19 eV, the removal of the strain results in a gap enlargement that strongly depends on the system considered, ranging from 0.07 eV (Si$_{35}$ NC) to 1.29 eV (Si$_{17}$ NC). This result reveals that the NCs can be subjected to amounts of strain that differs very much one from another. Besides, for strained NCs, also the substitution of -OH terminations with -H produces gap enlargements ranging from 0.3 eV (Si$_{17}$ NC) to 3.06 eV (Si$_{10}$ NC). These strong variations of the gap due to strain removal and termination type will become clearer in the following, when the relationship between strain and oxidation will be taken into account.\\
As already observed,\cite{guerra} even for the completely relaxed hydroxided systems we have a strong discontinuity of the gap values with the NC size. Instead, the removal of the strain in the hydrogenated systems produces a trend of the HOMO-LUMO gaps that follows the QC, as expected.\cite{seino}\\

\begin{figure}[b]\begin{center}
\centerline{\psfig{file=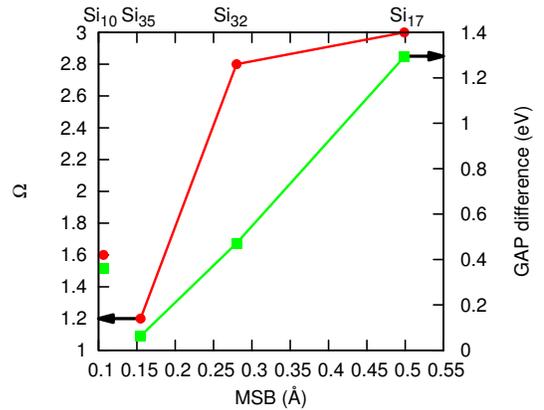,width=5.5cm,angle=270}}
\caption{(color online) Dependence on the MSB value for (squares) HOMO-LUMO gap difference between the hydroxided-relaxed and -strained NCs, and (circles) oxidation degree. The lines are drawn to guide the eye.}
\label{strain_gap}\end{center}\end{figure}

\begin{figure*}[t]\begin{center}
\centerline{\psfig{file=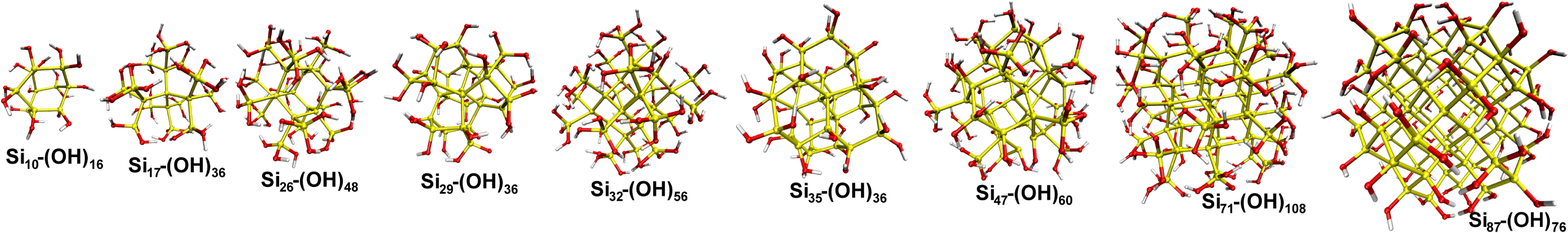,width=18cm}}
\caption{(color online) Stick and ball pictures of the final optimized set of hydroxided structures. Red (dark gray), white (light gray), and yellow (gray) spheres represent respectively the O, H, and Si atoms.}
\label{clusterset}\end{center}\end{figure*}

To analyze in detail the role of the embedding matrix, as for the contribution to the absorption spectrum, like as for the induced strain, we consider the reference case of the Si$_{32}$ NC. The absorption spectra [expressed by the imaginary part of the dielectric function, calculated within the DFT--random-phase approximation (RPA)] of Si$_{32}$/SiO$_2$, s-Si$_{32}$-OH, and optimized r-Si$_{32}$-OH are presented in Figure \ref{eps2_strain}. We note that the hydroxided-strained NC is able to reproduce very well the spectrum of the full Si/SiO$_2$ system in the 0-7 eV region, that is indeed associated to the NC+interface contribution.\cite{guerra,koponen} Instead, the removal of the strain produces an enlargement of the HOMO-LUMO gap,\cite{peng,franceschetti} and a consequent blue-shift of the absorption spectra in this region. For higher energies (above 7 eV), the spectra of the suspended passivated clusters do not match that of the composite system, due to the absence of the silica matrix. These are very general results, that we have verified in all the cases of Si/SiO$_2$ NCs. As already demonstrated,\cite{guerra,zhou,seino,wolkin,dohnalova} the substitution of hydroxyl groups with hydrogens results in a strong blue-shift of the absorption threshold, and the disappearance of the low-energy features that are related to the Si/SiO$_2$ interface.\\
To conclude this section, we observe that the considered NCs present different configurations of the Si/SiO$_2$ interface. In particular, the silicon atoms at the interface have a different number of nearby silica oxygens. Intuitively, we expect that a greater number of oxygen atoms  could exert a larger strain on the interface atoms. By defining $\Omega$ as the number of average oxygens per silicon at the interface, i.e. as the ratio between the number of oxygen atoms and the number of silicon atoms bonded to them, we could compare the amount of strain with the oxidation $\Omega$. By looking at Figure \ref{strain_gap} we observe, by comparing the properties of the hydroxided-relaxed and -strained NCs, that both the oxidation degree (upper curve) and the HOMO-LUMO gap shift (lower curve) generally increase with the strain, as expected. A discrepancy is present for the Si$_{10}$ NC, that could be reasonably addressed to the fact that this is a borderline case in which all the Si atoms of the NC are localized at the interface. Also, the discontinuity could arise from the contribution of the angular strain (not considered here), and from the non uniform behaviour of the strain distribution near the interface.\cite{yilmaz_bulutay_cagin}\\
At the end we summarize the effects of $\Omega$ on the energy gap by reporting, in Table \ref{tabella2}, the HOMO-LUMO gap-shift resulting from the comparison between the suspended-relaxed and -strained NCs, with the corresponding $\Omega$ value. As expected, $\Omega$ and the gap-shift present the same trend.

\begin{table}[h!]\begin{center}\begin{tabular}[t]{c @{\hspace{0.6cm}} c
@{\hspace{0.3cm}} c @{\hspace{0.3cm}} c @{\hspace{0.3cm}} c}
\toprule
               & Si$_{35}$-OH & Si$_{10}$-OH & Si$_{32}$-OH & Si$_{17}$-OH   \\
\midrule
    gap-shift (eV)  & 0.07	& 0.37       & 0.48      & 1.29       \\
    $\Omega$        & 1.2	& 1.6        & 2.8       & 3.0        \\

\bottomrule
\end{tabular}
\caption{Oxydation degree, $\Omega$, and HOMO-LUMO gap-shift resulting from the comparison between the suspended-relaxed and -strained NCs.}
\label{tabella2}\end{center}\end{table}

\section{Oxidation and Strain}\label{sec_strainox}

\begin{figure}[b]\begin{center}
\centerline{\psfig{file=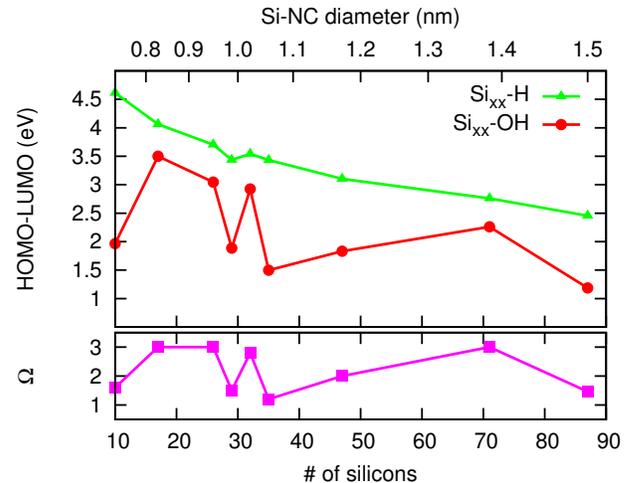,width=6.5cm,angle=270}}
\caption{(color online) HOMO-LUMO gaps for the hydroxided (circles)
and hydrogenated (triangles) NCs, together with the
oxidation/hydrogenation degree $\Omega$ (squares).}
\label{strainox}\end{center}\end{figure}

In this section we explore the possibility that the strong discontinuity in the energy gap values for the NCs analyzed above, could be related to the different passivation regime of the interface silicon atoms. It is worth noting that a recent size-dependent experimental study of Si 2$p$ core-level shift at Si-NC/SiO$_2$ interface showed that the shell region around the Si-NC bordered by SiO$_2$ consists of the three Si suboxide states, Si$^{1+}$, Si$^{2+}$, and Si$^{3+}$, where for NCs smaller than about 3 nm the presence of suboxides with higher oxidation becomes favoured.\cite{kimkim} In order to focus on this aspect, we produced a set of suspended NCs, each with different geometry, size, and oxidation degree (see Fig. \ref{clusterset}). The strain has been intentionally removed from all the NCs, through geometry optimization, to elucidate further the correlation between the final gap value and $\Omega$. In parallel to the hydroxided NCs we also built the corresponding hydrogenated systems, and we then calculated the HOMO-LUMO gaps for all the structures (see Fig. \ref{strainox}).
What emerges, for the hydroxided case, is a strong and clear correlation between $\Omega$ and the gap value. For these sizes, the dependence of the gap on $\Omega$ is much stronger than that due to QC, so that a small variation in size could correspond to a large variation of the gap value. The effects of QC are still alive, but strongly modulated by that of oxidation. As already anticipated, the hydrogenated case does not instead depend on $\Omega$ but simply follows the QC. As suggested by other works\cite{vasiliev,zhou}, we expect that for large clusters the gap becomes independent on the passivation regime, just following the QC.\\
Considering the results from the previous Section, we could guess that the inclusion of the strain would reduce the HOMO-LUMO gaps of Fig. \ref{strainox}, depending (non linearly) on the $\Omega$ value. In our opinion, this would reduce the gap-fluctuations amplitude, and also influence the QC trend. In fact, by looking at Table \ref{tabella1} we see that, by comparing for example the Si$_{17}$ and the Si$_{32}$ embedded NCs, the final gap do not follow the QC rule, even if they present similar oxydation degrees. In this case the strain completely overrules the oxidation effect. Instead, for very different $\Omega$ values the final gap still present a clear correlation to the oxidation degree, again undependently on the NC size. These considerations suggest a final picture in which the gap of small embedded NCs show an oscillating behaviour with size, and not strictly following the QC, as already observed in other works.\cite{koponen,zhou}
The large broadening of the PL spectra, observed also at low temperatures, \cite{dovrat,klimov,delerue,lin_chen,hernandez,kanemitsu2,ray_green,brongersma,rolver,averboukh,nayfeh,gardelis} even for apparently monodispersed multilayered samples,\cite{cazzanelli} could be a possible consequence of such fluctuations of the electronic gap with the NC size.\\

The case of interface defects has been deeply investigated, sometimes suggesting that they have a small influence on the radiative recombination,\cite{ramos_bechstedt,delerue,derr_rosei,hill_whaley} and other times being indicated as the primary source for radiative emission.\cite{godefroo,averboukh,ray_green,kanemitsu2,kanemitsu,iwayama} Moreover, the presence of other oxygen-bond types is frequently reported in experimental samples, beside the single-bond type discussed here. Several works report the Si-O-Si bridge bond as the most stable configuration, and the Si=O double bond as the less energetically favoured.\cite{nguyen,jurbergs,seino_bechstedt,galli} Besides, these alternative bond types are especially favoured for large clusters sizes.\cite{kelires1,colombo} In conclusion, we believe that these additional parameters (defects, bond types) would not modify the general picture related to the parameter $\Omega$ discussed above, making it a crucial quantity for the characterization of small Si-NCs.

\section{Conclusions}\label{sec_conclusions}
Considering the $\beta$-cristobalite phase of the SiO2 we have generated silicon nanocrystals of different size which are embedded in a silica matrix. The relaxed structures show that the surrounding matrix always produces some strain on the nanocrystal, especially at the Si/SiO$_2$ interface. By including the strain on the suspended hydroxided nanocrystals we were able to reproduce the electronic and optical properties of the full Si/SiO$_2$ systems. Besides, the removal of such strain by a total relaxation will produce a significant reduction of the electronic gap. We found evidences that the amount of strain exerted on the nanocrystal is connected to the interface configuration, in particular to the number of oxygens per interface silicons (oxidation degree). Moreover
we revealed that, while the quantum confinement dominates in the hydrogenated nanocrystals of all sizes, the behaviour of hydroxided and embedded nanocrystals show a strong correlation between the final value of the electronic gap and the oxidation degree, that seems to override the effects of quantum confinement for nanocrystals of diameters below 2 nm. The final picture indicates a competition between the oxidation and the strain, in which the former tends to blue-shift the absorption spectra, and the latter to red-shift it.
At the end we suggest a final picture in which the gap of small embedded NCs shows an oscillating behaviour with size, not strictly following the QC rule.
The strong fluctuations of the electronic gap with the nanocrystal size could be responsible for the large broadening of the PL spectra that have been observed experimentally, also at low temperatures.\cite{dovrat,klimov,delerue,lin_chen,hernandez,kanemitsu2,ray_green,brongersma,rolver,averboukh,nayfeh,gardelis,cazzanelli}\\

\subsection*{ACKNOWLEDGEMENTS}
This work is supported by PRIN2007 and MAE-CNR Italia-Turchia. We acknowledge CINECA CPU time granted by INFM (Progetto Calcolo Parallelo).


\begin{thebibliography}{}
\bibitem{bisi} O. Bisi, S. Ossicini, L. Pavesi, Surf. Sci. Rep. {\bf 38}, 5 (2000).
\bibitem{pavlockwood}L. Pavesi and D.J. Lockwood, Eds., Silicon Photonics, Springer, Berlin, 2004.
\bibitem{brongersma} M. L. Brongersma, P. G. Kik, A. Polman, Appl. Phys. Lett. {\bf 76}, 351-353 (2000).
\bibitem{iwayama} T. Shimitsu-Iwayama, T. Hama, D. E. Hole, I. W. Boyd, Solid-State Electronics {\bf 45}, 1487-1494 (2001).
\bibitem{hernandez} A. V. Hernandez, T. V. Torchynska, Y. Matsumoto, S. J. Sandoval, M. Dybiec, S. Ostapenko, L. V. Shcherbina, Microelectronics Journal {\bf 36}, 510-513 (2005).
\bibitem{rolver} R. Rolver, M. F\"{o}rst, O. Winkler, B. Spangenberg, H. Kurz, J. Vac. Sci. Technol. A {\bf 24}, 141-145 (2006).
\bibitem{godefroo} S. Godefroo, M. Hayne, M. Jivanescu, A. Stesmans, M. Zacharias, O. I. Lebedev, G. V. Tendeloo, V. V. Moshchalkov, Nature Nanotech. {\bf 3}, 174 (2008).
\bibitem{gardelis} S. Gardelis, A. G. Nassiopoulou, N. Vouroutzis, N. Frangis, J. Appl. Phys. {\bf 105}, 113509 (2009).
\bibitem{delerue} C. Delerue, G. Allan, C. Reynaud, O. Guillois, G. Ledoux, F. Huisken, Phys. Rev. B {\bf 73}, 235318 (2006).
\bibitem{kanemitsu} Y. Kanemitsu, Thin Solid Films {\bf 276}, 44-46 (1996).
\bibitem{sychugov_linnros} I. Sychugov, R. Juhasz, J. Valenta, J. Linnros, Phys. Rev. Lett. {\bf 94}, 087405 (2005).
\bibitem{averboukh} B. Averboukh, R. Huber, K. W. Cheah, Y. R. Shen, G. G. Qin, Z. C. Ma, W. H. Zong, J. Appl. Phys {\bf 92}, 3564-3568 (2002).
\bibitem{antonova} I. V. Antonova, M. Gulyaev, E. Savir, J. Jedrzejewsky, I. Balberg, Phys. Rev. B  {\bf 77} 125318 (2009).
\bibitem{salivati} N. Salivati, N. Shuall, E. Baskin, V. Garber, J. M. McCrate, J. G. Ekerdt, J. Appl. Phys. {\bf 106}, 063121 (2009).
\bibitem{liptak} R. W. Liptak, U. Kortshagen, S. A. Campbell, J. Appl. Phys. {\bf 106}, 064313 (2009).
\bibitem{coxon} P. R. Coxon, Y. Chao, B. R. Horrocks, M. Gass, U. Bangert, L. \v{S}iller, J. Appl. Phys. {\bf 104}, 084318 (2008).
\bibitem{konig} D. K\"onig, J. Rudd, M. A. Green, G. Conibeer, Phys. Rev. B {\bf 78}, 035339 (2008).
\bibitem{lehtonen} O. Lehtonen, D. Sundholm, Phys. Rev. B {\bf 74}, 045433 (2006).
\bibitem{dalnegro} L. Sirleto, M. A. Ferrara, I. Rendina, S. N. Basu, J. Warga, R. Li, L. Dal Negro, Appl. Phys. Lett. {\bf 93}, 251104 (2008).
\bibitem{empedocles_norris_bawendi} S. A. Empedocles, D. J. Norris, M. G. Bawendi, Phys. Rev. Lett. {\bf 77}, 3873 (1996).
\bibitem{credo} G. M. Credo, M. D. Mason, S. K. Buratto, Appl. Phys. Lett. {\bf 74}, 1978 (1999).
\bibitem{guerra} R. Guerra, I. Marri, R. Magri, L. Martin-Samos, O. Pulci, E. Degoli, S. Ossicini, Phys. Rev. B {\bf 79}, 155320 (2009).
\bibitem{vasiliev} I. Vasiliev, J. R. Chelikowsky, R. M. Martin, Phys. Rev. B {\bf 65}, 121302(R) (2002).
\bibitem{allan_delerue_lannoo} G. Allan, C. Delerue, M. Lannoo, Phys. Rev. Lett. {\bf 76}, 2961 (1996).
\bibitem{BC} H. Kageshima, K. Shiraishi, in: M. Scheffler, R. Zimmermann (Eds.), Proc. 23rd Int. Conf. Phys. Semicon., World Scientific, Singapore, p. 903, (1996).
\bibitem{mluppi1} M. Luppi, S. Ossicini,  J. Appl. Phys., {\bf 94}, 2130 (2003).
\bibitem{mluppi2} M. Luppi, S. Ossicini, Phys. Rev. B {\bf 71}, 035340 (2005).
\bibitem{flyura} F. Djurabekova, K. Nordlund, Phys. Rev. B {\bf 77}, 115325 (2008).
\bibitem{siesta1} P. Ordej\'{o}n, E. Artacho, J. M. Soler, Phys. Rev. B (Rapid Comm.) {\bf 53}, R10441 (1996).
\bibitem{siesta2} J. M. Soler, E. Artacho, J. D. Gale, A. Garc\'{\i}a, J. Junquera, P. Ordej\'{o}n, D. S\'{a}nchez-Portal, J. Phys.: Condens. Matt. {\bf 14}, 2745 (2002).
\bibitem{espresso} P. Giannozzi, S. Baroni, N. Bonini, M. Calandra, R. Car, C. Cavazzoni, D. Ceresoli, G. L. Chiarotti, M. Cococcioni, I. Dabo, A. Dal Corso, S. Fabris, G. Fratesi, S. de Gironcoli, R. Gebauer, U. Gerstmann, C. Gougoussis, A. Kokalj, M. Lazzeri, L. Martin-Samos, N. Marzari, F. Mauri, R. Mazzarello, S. Paolini, A. Pasquarello, L. Paulatto, C. Sbraccia, S. Scandolo, G. Sclauzero, A. P. Seitsonen, A. Smogunov, P. Umari, R. M. Wentzcovitch, J. Phys.: Condens. Matt. {\bf 21}, 395502 (2009) \url{http://dx.doi.org/10.1088/0953-8984/21/39/395502}.
\bibitem{bulutay} D. E. Yilmaz, C. Bulutay, T. \c{C}a\u{g}in, Phys. Rev. B \textbf{77}, 155306 (2008).
\bibitem{kroll} P. Kroll, H. J. Schulte, Phys. Stat. Sol. B {\bf 243} (2006).
\bibitem{koponen} L. Koponen, L. O. Tunturivuori, M. J. Puska, R. M. Nieminen, Phys. Rev. B {\bf 79}, 235332 (2009).
\bibitem{seino} K. Seino, F. Bechstedt, P. Kroll, Nanotechnology {\bf 20}, 135702 (2009).
\bibitem{peng} X.-H. Peng, S. Ganti, A. Alizadeh, P. Sharma, S. K. Kumar, S. K. Nayak, Phys. Rev. B {\bf 74}, 035339 (2006).
\bibitem{franceschetti} A. Franceschetti, Phys. Rev. B {\bf 76}, 161301(R) (2007).
\bibitem{zhou} Z. Zhou, L. Brus, R. Friesner, Nano Lett. {\bf 3}, 163 (2003).
\bibitem{wolkin} M. V. Wolkin, J. Jorne, P. M. Fauchet, G. Allan, C. Delerue, Phys. Rev. Lett. {\bf 82}, 000197 (1999).
\bibitem{dohnalova} K. Dohnalov\'{a}, K. K\r{u}sov\'{a}, I. Pelant, Appl. Phys. Lett. {\bf 94}, 211903 (2009).
\bibitem{yilmaz_bulutay_cagin} D. E. Yilmaz, C. Bulutay, T. \c{C}a\u{g}in, Appl. Phys. Lett. {\bf 94}, 191914 (2009).
\bibitem{kimkim} S. Kim, M. C. Kim, S-H. Choi, K. J. Kim, H. N. Hwang, C. C. Hwang, Appl. Phys. Lett. \textbf{91}, 103113 (2007).
\bibitem{dovrat} M. Dovrat, Y. Shalibo, N. Arad, I. Popov, S.-T. Lee, A. Sa'ar, Phys. Rev. B {\bf 79}, 125306 (2009).
\bibitem{klimov} M. Sykora, L. Mangolini, R. D. Schaller, U. Kortshagen, D. Jurbergs, V. I. Klimov, Phys. Rev. Lett. {\bf 100}, 067401 (2008).
\bibitem{lin_chen} S.-W. Lin, D.-H. Chen, Small {\bf 5}, 72-76 (2009).
\bibitem{kanemitsu2} Y. Kanemitsu, Y. Fukunishi, M. Iiboshi, S. Okamoto, T. Kushida, Physica E {\bf 7}, 456-460 (2000).
\bibitem{ray_green} M. Ray, K. Jana, N. R. Bandyopadhyay, S. M. Hossain, D. Navarro-Urri\'{o}s, P. P. Chattyopadhyay, M. A. Green, Solid State Communications {\bf 149}, 352-356 (2009).
\bibitem{nayfeh} G. Belomoin, J. Therrien, M. Nayfeh, Appl. Phys. Lett. {\bf 77}, 779-781 (2000).
\bibitem{cazzanelli} M. Cazzanelli, D. Navarro-Urri\'{o}s, F. Riboli, et al., J. Appl. Phys. {\bf 96}, 3164-3171 (2004).
\bibitem{ramos_bechstedt} L. E. Ramos, J. Furthm\"{u}ller, F. Bechstedt, Phys. Rev. B {\bf 71}, 035328 (2005).
\bibitem{derr_rosei} J. Derr, K. Dunn, D. Riabinina, F. Martin, M. Chacker, F. Rosei, Physica E {\bf 41} 668-670 (2009).
\bibitem{hill_whaley} N. A. Hill, K. B. Whaley, Phys. Rev. Lett. {\bf 75}, 1130 (1995).
\bibitem{nguyen} N. B. Nguyen, C. Dufour, S. Petit, J. Phys. Cond. Matt. {\bf 20}, 455209 (2008).
\bibitem{jurbergs} D. Jurbergs, E. Rogojina, L. Mangolini, U. Kortshagen, Appl. Phys. Lett. {\bf 88}, 233116 (2006).
\bibitem{seino_bechstedt} K. Seino, J.-M. Wagner, F. Bechstedt, Appl. Phys. Lett. {\bf 90}, 253109 (2007).
\bibitem{galli} A. Puzder, A. J. Williamson, J. C. Grossman, G. Galli, J. Am. Chem. Soc. {\bf 125}, 2786-2791 (2003).
\bibitem{kelires1} G. Hadjisavvas, P.C. Kelires, Physica E {\bf 38}, 99 (2007).
\bibitem{colombo} M. Ippolito, S. Meloni, L. Colombo, Appl. Phys. Lett. {\bf 93}, 153109 (2008).
\end{thebibliography}
\end{document}